# Physical Constraints for the Stoneham Model for Light-Dependent Magnetoreception

J. Espigulé-Pons, C. Goetz, A. Vaziri, M. Arndt
University of Vienna, QuNaBioS, Boltzmanngasse 5 and Dr. Bohr-Gasse 7, A-1090 Vienna, Austria

ABSTRACT    A new biophysical model for magnetoreception in migratory birds has recently been proposed (Stoneham et al. 2012. *Biophys. J.* 102: 961–968). In this photo-induced radical pair (RP) model the signal transduction mechanism was physical rather than chemical in nature, as otherwise generally assumed in the literature. The proposal contains a magnetosensor and a signal transduction mechanism. The sensor would be an electric dipole related to a long lived triplet state of an RP. This makes it sensitive to the geomagnetic field via the Zeeman interaction. The field of the electric dipole moment would then promote isomerization from *cis*-to-*trans* in the retinal of a nearby rhodopsin. This would trigger the neuronal signal. Here we gather several observations from different works that constrain the feasibility of this physical model. In particular we argue that the perturbation of rhodopsin by a local electric field from a nearby electric dipole ($10^6$ V/m) cannot modify the field in the binding pocket of rhodopsin ($10^9$ V/m) sufficiently to trigger the isomerization of cis-retinal. The dipole field is much weaker than those from other sources in the vicinity which are known not to promote isomerization.

Address reprint requests and inquiries to jofre.espigule@univie.ac.at

## INTRODUCTION

Magnetoreception is the ability of living organisms to sense magnetic fields. Despite the increasing number of species reported to have this extra sense (1,2), the biophysical mechanisms behind this ability are still obscure. The models for magnetoreception can be divided into two main groups: magnetite-particle-based (3,4) and radical-pair-based models (5–11). Some authors have also combined both (12,13) or have exploited the idea of electromagnetic induction in elasmobranch fishes, such as sharks (1).

A recent hypothesis by Stoneham et al. (7) has attracted interest (14,15) as an explanation for of the magnetic signal transduction in migratory birds (16–18), where behavioral experiments favor radical-pair-based models. While many authors have assumed the signal transduction to be of chemical nature (6,16) a physical signal transduction is *a priori* equally conceivable. Stoneham et al. (7) suggested that an electric dipole may originate from a radical pair (RP) and that the electric field of the dipole could enhance the isomerization of *cis*-retinal in a nearby rhodopsin, which would then trigger a neuronal signal (Fig. 1).

This model can account for many observations, such as a wavelength-dependent inclination compass in the retina and the disruption of the compass by weak radiofrequency fields (19,20,21). In spite of that, there are strong arguments that question the feasibility of the model, which we outline below.

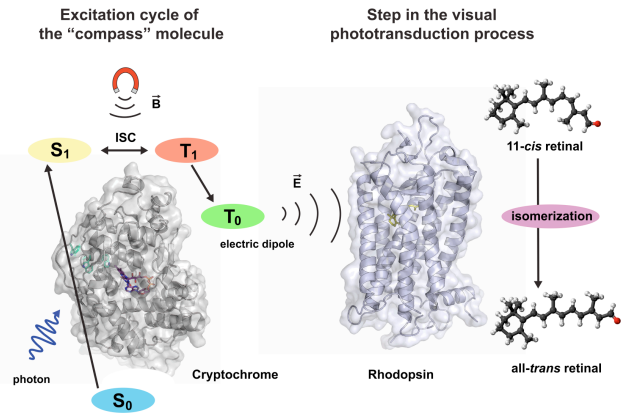

**Figure 1:** Stoneham model (7): The first step consists of the absorption of a blue/green photon by the "compass molecule". This is motivated by behavioural experiments with migratory birds. The protein Cryptochrome is the main candidate as a photosensor. After the absorption process a radical pair is formed in the singlet state $S_1$. It can coherently evolve to the triplet $T_1$ under the joint influence of hyperfine and Zeeman interactions. This process is described here as intersystem crossing (ISC). The population of the triplet state may vary as a function of the protein orientation to the external magnetic field. Finally, a long lived triplet $T_0$ state, populated via $T_1$ creates an electric dipole whose field gradient is proposed to promote isomerization of 11-*cis* retinal in a nearby rhodopsin. This would then trigger a neuronal signal in the visual system. Note, that $T_0$ designates a low energy triplet instead of the triplet zero in the multiplet $T_1$. (Protein structures are obtained from PDB, Figures in PYMOL).

**Caveat 1. Electric field induced isomerization**

Stoneham et al. (7) proposed the field-induced *cis*-retinal isomerization in rhodopsin to be similar to that of azobenzene derivatives in solution, which was observed earlier (22,23). Electric fields between $10^4$-$10^6$ V/m were shown to induce isomerization of azobenzene in the presence of electron transfer from a nearby cathode to the *cis*-conformer, as indicated in Fig. 2 (22). This cannot be simply extrapolated to rhodopsins.

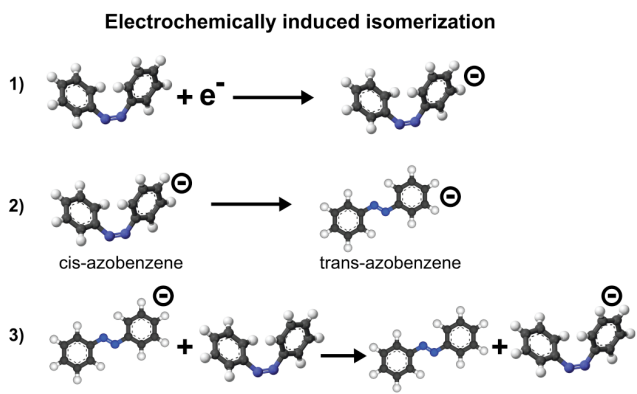

**Figure 2:** Electrochemical reaction routes to isomerization of *cis*-azobenzene. 1) Electrolysis induced near the cathode by an applied electric field creates free electrons that reduce *cis*-azobenzene isomers. 2) The energy barrier between charged *cis*-azobenzene and *trans*-azobenzene is lowered enough to enable the isomerization to *trans*-azobenzene. 3) Finally, a neutral *trans*-azobenzene is obtained via electron transfer to another *cis*-isomer.

Although STM experiments were able to show electric field-induced isomerization even without electron transfer (24, 25), the required local field was of the order of $10^9$-$10^{10}$ V/m. The azobenzene was bound to a surface in ultra-high vacuum at low temperatures and the effect was only observed on Au(111) but not on Cu(111) and Au(100) surfaces (26).

Recently, the field-induced isomerization of a biomolecule has also been observed, namely in the enzyme cyclophilin A, which catalyzes proline isomerization by an electrostatic handle mechanism (27). The electric field of about $10^{10}$ V/m in the active site of cyclophilin A, is, however, four orders of magnitude higher than proposed by Stoneham et al (7). Such high fields can reduce the potential energy barrier between the *cis* and the *trans* state, resulting in a speed-up of the isomerization process by four to five orders of magnitude, i.e. from minutes to milliseconds.

Stoneham et al. (7) also cited studies that showed electric field effects on bacteriorhodopsin. These works did, however, not report on isomerization of retinal by an external electric field (28) but on migration and orientation of negatively charged proteins (29). Another field effect reported for bacteriorhodopsin was the modification of the photocycle time (28,29).

Finally, field-related changes of the retinal absorption spectrum could indirectly affect photo-isomerization and thus carry magnetic field information. This last case was not explicitly considered by Stoneham et al. (7) but it would require the absorption of a second photon in the retinal immediately after the formation of the electric dipole has been formed via photoabsorption in the nearby cryptochrome.

The intensity needed for the coupled cryptochrome-rhodopsin system to absorb two photons within the life-time of the putative RP and charge-separated triplet $T_0$ state ($t_{RP}$ = 1 ms) is estimated to be:

$$\frac{2E_\gamma}{t_{RP} \cdot \sigma_{abs}} = 10 \, W/m^2 \qquad (1)$$

For green light ($\lambda$ = 555 nm, $E_\gamma$ = 2.23 eV) and a typical absorption cross section of $\sigma_{abs}$ = $5\times10^{-17}$ cm$^2$ this corresponds to 6800 lux. That is more than six orders of magnitude larger than the value needed by migratory birds to navigate in a moonless night, where the light may be as weak as 0.002 lux (11,30).

**Caveat 2. Fields in the binding pocket of rhodopsin**

Even if one accepts the possibility of *cis*-retinal isomerization by electric fields, recent simulations and experimental evidence show that the electrostatic potential in the binding pocket of rhodopsin gives rise to electric fields of about $10^9$ V/m across the retinal molecule (31–33) (see Fig. 3). Small conformational changes of rhodopsin, for example induced by thermal fluctuations, would already produce E-field perturbations that can easily exceed those caused by the RP dipole.

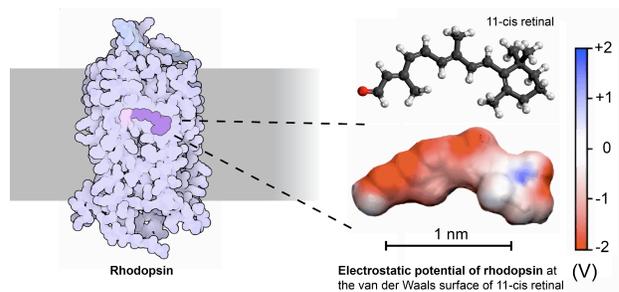

**Figure 3:** Simulation of the electrostatic potential of rhodopsin at the van der Waals surface of 11-*cis* retinal. The Figure is composed from a rhodopsin protein image taken from RCSB PDB (D. Goodsell), 11-*cis* retinal structure image created with DSV and electrostatic potential image adapted from (31).

The field created by two elementary charges that separated from each other by 1 nm and from the rhodopsin-bound retinal by 10 nm amount to only:

$$E = \frac{e}{4\pi\varepsilon_0\varepsilon_r}\sqrt{\left(\frac{x}{r_+^3} - \frac{x}{r_-^3}\right)^2 + \left(\frac{y}{r_+^3} - \frac{y}{r_-^3}\right)^2} \approx 10^5 V/m \quad (2)$$

Since it is known that 11-*cis* retinal is relatively stable under thermal fluctuations (34–36) we conclude that the local electric field from the nearby electric dipole is too weak to isomerize 11-*cis* retinal.

**Caveat 3. Electric properties of the solvent**

Both the sensor and transductor molecule are in contact with the cytosol. Water is a polar solvent and its dielectric constant at room temperature ($\varepsilon_r$=80) is high in comparison to that of the protein ($\varepsilon_r$ =2-4) (37) in which the radical pair is formed. The field in Eq. (2) is therefore about 80 times smaller than estimated by Stoneham et al. (7), who neglected solvent effects.

A measure for the interaction between charge particles in solution is the *Bjerrum* length $\lambda_B$. It defines the separation between two elementary charges at which the electrostatic potential becomes comparable to the thermal energy $k_BT$:

$$\lambda_B = \frac{e^2}{4\pi\varepsilon_0\varepsilon_r k_B T} \approx 0.7 nm \quad (3)$$

Since the electric field of the RP dipole is certainly smaller than that of an elementary charge alone, we can exclude induced isomerization by an RP dipole at 10 nm.

Furthermore, the concentration of $K^+$ ions in the cytosol may amount to about 155 mM (38). Correspondingly, there are about 100 potassium ions in a solvent element of $10\times10\times10$ nm$^3$ size. They would shield the effective field of the RP dipole and cause competing time-dependent fields when drifting around the protein.

**Caveat 4. The ubiquity of charges in proteins**

Charges in proteins are ubiquitous (39,40). In particular plant cryptochrome has positively charged residues on the protein surface (41). Because of their similar structure animal cryptochromes should also expose charged residues. Also bovine rhodopsin is known to have two charged residues in addition to the protonated Schiff base (42). These observations indicate that other electric field sources around retinal will surpass the field of the RP dipole, substantially.

**Caveat 5. The elusive long-lived triplet state, $T_0$**

The existence of a long-lived charge-separated triplet $T_0$ in cryptochrome has not been observed, so far.

First, if $T_0$ is required to generate a significant electric field, then it must have positive and negative charges separated by 1 nm or more. As such, hyperfine interactions would be expected to cause coherent triplet-singlet intersystem crossing on a sub-microsecond timescale, as in the case of the FAD-Trp radical pair. One would then have to assume that both singlet and triplet states of this radical pair were long-lived and that neither of the radicals became protonated or deprotonated – as happens for both the FAD and Trp radicals (16).

Second, the existence of $T_0$ is no better an explanation for any disorienting effect of radiofrequency fields than is the FAD-Trp radical pair ($S_1$-$T_1$) itself. In both cases, astonishingly slow (>100 μs) spin relaxation would be required to allow a 15 nT RF field to have a significant effect (8,19,43).

In conclusion, it seems highly unlikely that a $T_0$ species could be formed in cryptochrome. The triplet states of both flavin and tryptophan are significantly higher in energy than the radical pair. It seems very improbable that the FAD-Trp radical pair could populate triplet states of anything else in its vicinity.

**Colocalization and partial orientation of cryptochrome close to rhodopsin?**

Nevertheless cryptochrome is the main candidate for a magnetosensitive molecule (44–47). In order for the bird to be sensitive to the direction of the magnetic field, the compass molecules must be immobilized at least partially (6, 48, 49). Does cryptochrome bind to a membrane protein or tether to the membrane itself? If it does and interacts with rhodopsin we propose that a different transduction mechanism might explain magnetoreception, as well.

A direct protein-protein interaction, via electrostatic potentials changes, may induce a conformational change in rhodopsin (50), which triggers isomerization. In this case charged residues in cryptochrome would be the main drivers. This hypothesis could be partially tested using a fluorescence complementation technique, where two fluorescent protein fragments would be attached, one to cryptochrome and the other to rhodopsin. This can cause fluorescence when the proteins interact with each other (51). The emerging light would prove the colocalization of the sensor and transductor molecules.

At present, this idea cannot be implemented in migratory birds, since the required genetic tools are not yet developed. But antibodies can be made in order to tackle this question. Additionally the recent emerging CRISPR/CAS9 genome editing technology (52) may open up new paths for genetic manipulation of non-standard model organisms. Previous studies have already shown the localization of different cryptochromes in the outer layers of the retina (43,53,54), in particular also the cones (18) of migratory birds. A proof of sub-cellular colocalization of cryptochrome and rhodopsin is, however, not yet available.

As an alternative to the electric signal transduction a chemical pathway has been proposed (9,10) to rely on the formation of superoxide ($O_2^-$) in the excited cryptochrome. It has, however, been argued that this is unlikely (55,56).

Both a chemical and a physical inclination compass would still require the partial immobilization of the compass proteins. This could possibly be achieved by constraining the diffusion in the outer segment (*OS*) of the rod cells (57,58), as shown in Fig. 4. For green fluorescent protein (GFP) a diffusion constant of $D_{OS} = 0.079 \pm 0.009$ m$^2$/s was found in this segment (59). This is 1000 times smaller than in aqueous solution. Since cryptochrome is comparable in size and since its surface has hydrophobic surfaces patches, a partial orientation within the hydrophobic membrane of the rod disks appears plausible. In a practical test, the signal to noise could be enhanced by increasing the number of cryptochromes expressed in the rods or cones cells. This is in accordance with recent experimental observations which have showed that migratory birds have high levels of cryptochrome in the retina during night, when they are migrating, while the level of cryptochromes decreases during night in non-migration periods and non-migrating species (18,43,60).

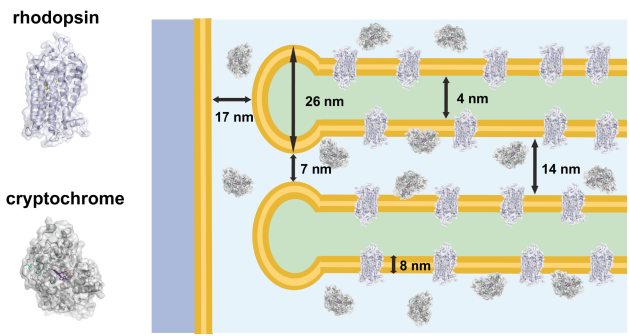

**Figure 4:** Structure of the outer segment of a rod cells. (Cryptochrome has a diameter of 7 *nm* approximately). The protein structures are obtained from PDB, Figures in PYMOL and image of *OS* has been adapted from (54).

Summarizing, while the electric dipole moment of a photo-induced radical pair (7), seems insufficient to explain magnetoreception of migratory birds, the original proposal still inspires future explorations of the role of the rhodopsin protein as a transducer for a cryptochrome compass.

## ACKNOWLEDGEMENT

We acknowledge Professor W.H. Koppenol for pointing to the ubiquity of charges in proteins. We gratefully acknowledge fruitful comments by Professor P.J. Hore.

We acknowledge financial support by the European Research Council in project 304886.